# New opportunities for integrated microwave photonics

David Marpaung[1], Jianping Yao[2], and Jose Capmany[3]

[1]Laser Physics and Nonlinear Optics Group, Faculty of Science and Technology, MESA+ Institute for Nanotechnology, University of Twente, the Netherlands

[2]School of Electrical Engineering and Computer Science, University of Ottawa, Canada

[3]ITEAM Research Institute, Universitat Politècnica de València, Spain

Recent advances in photonic integration have propelled microwave photonic technologies to new heights. The ability to interface hybrid material platforms to enhance light-matter interactions has led to the developments of ultra-small and high-bandwidth electro-optic modulators, frequency-synthesizers with the lowest noise, and chip signal processors with orders-of-magnitude enhanced spectral resolution. On the other hand, the maturity of high-volume semiconductor processing has finally enabled the complete integration of light sources, modulators, and detectors in a single microwave photonic processor chip and has ushered the creation of a complex signal processor with multi-functionality and reconfigurability similar to their electronic counterparts. Here we review these recent advances and discuss the impact of these new frontiers for short and long term applications in communications and information processing. We also take a look at the future perspectives in the intersection of integrated microwave photonics with other fields including quantum and neuromorphic photonics.

## Introduction

The use of optical devices and techniques to generate, manipulate, transport, and measure high speed radiofrequency signals, widely known as microwave photonics (MWP), has been the focus of intense research activities in recent years [1-5]. The promise of abundant processing bandwidth obtained from upconverting the radiofrequencies to the optical frequencies, the availability of low loss optical fibers as transport medium, and the flexibility in tailoring the radiofrequency response over decades of frequency unlike anything achievable by traditional RF systems have been cited as key drivers in the early development of the technology. Soon after, these promises were delivered, with landmark demonstrations. Among others, generation of ultra-broadband signals [6], distribution and transport of RF over fiber [7], wideband tunable filters [8], and photonics-enhanced radar system [9] ensued. These progresses subsequently positioned MWP as a prime technology solution to the impending challenges in communications including the bandwidth bottleneck in communication systems [10] and the internet of things, provided that another hurdle is overcome, namely size, reliability, and cost effectiveness. While impressive, those landmark results were demonstrated in bulky systems composed of relatively expensive discrete fiber-optic components, which are sensitive to external perturbations such as vibrations and temperature gradients.

It is thus serendipitous, but truly essential, that the rise of MWP technology was paralleled by the surge of photonic integration technologies. The convergence of the two fields, aptly termed integrated microwave photonics, soon followed with profound impacts [5, 11]. Leveraging photonic integration allowed dramatic footprint reduction of MWP systems with fairly high complexity, making it more comparable to RF circuits. But this advantage often comes at the expense of optical loss resulted from scattering losses in the optical waveguides as well as from fiber-to-chip coupling, which is needed as in the majority of the early demonstrations, the light sources, modulators, and detectors were located off-chip. Optical loss in this case is important because an increase in this parameter translates quadratically into the RF loss in the circuit [5]. For these reasons, the efforts in this initial stage of integrated microwave photonics were focused on reducing on-chip losses [12], integrating as many components as possible in a single chip [13], and to demonstrate device reconfiguration [14,15].

But integrated photonics offers much more than footprint reduction and complexity. For example, confining light in the small mode volume enhances its interaction with matter, most of the time through nonlinear optical processes, which created new technological tools for integrated MWP, such as Kerr microresonator combs [16], hybrid-organic plasmonic modulators [17], and on-chip stimulated Brillouin scattering [18]. These tools, married with entirely new concepts, for example making a universal reconfigurable processor [19], significantly alter the capabilities of microwave photonic systems to achieve higher performance, including modulation bandwidth, spectral resolution, noise performance, and reconfigurability. On the other hand, major advances in chip integration using a single material (monolithic) or multi-materials (hybrid or heterogeneous) have allowed the integration of all the key components of integrated MWP systems in a single chip [20]. The synergies of integration, advanced functionalities, and high performance have been the key highlights of the field in recent times.

Here, we review the most recent advances in integrated MWP. We will briefly cover the material platforms and functionalities in the field, in a more traditional sense, but will refer the reader to a more comprehensive review on this topic [5]. Instead, here we focus on the new technological tools for integrated microwave photonics, derived from recent breakthroughs and advances in integrated optics. These new advances considerably expanded the performance and the scope of the field, allowing intersections with emerging technologies such as quantum photonics, optomechanics, and neuromorphic photonics. We offer the perspective of how these fields can interact as well as the short and long-term applications of the technology.

**Major material platforms**

At its inception [5], integrated MWP was mainly demonstrated in rather diverse material platforms that include GaAs, lithium niobate, and doped silica. But for the last 10 years, the majority of integrated MWP circuits were based on three key platforms for monolithic integration: Indium Phosphide (InP), silicon-on-insulator (SOI), and silicon nitride ($Si_3N_4$) (Fig 1(a-c)). Maturity in the fabrication process of these materials and their availability through cost-sharing initiatives that dramatically reduce the fabrication cost were two key drivers of this polarisation. Each of these materials have their own strengths and weaknesses, notwithstanding, impressive functions have been shown in the past several years.

InP [21] is the only material enabling the monolithic integration of various active and passive photonic components including lasers, modulators, optical amplifiers, tunable devices, and photodetectors and the platform allows creation of compact circuits with bending radius of the order if 100 microns. But optical waveguides in this material have relatively high losses of the order of 1.5 -3 dB/cm [22].

Silicon photonics [23], on the other hand, offers the compatibility with microelectronic Complementary Metal Oxide Semiconductor (CMOS) fabrication processes making electronic-photonic co-integration a real possibility. Silicon on Insulator (SOI) waveguides can exhibit a wide range of losses (0.1-3 dB/cm) and minimum bending radii (5-100 microns), depending on the thickness of the silicon layer ("thin" ~220 nm or "thick" ~3 microns) and the geometry of the waveguides (strip or rib). Advanced silicon circuits that use multiple waveguide structures to simultaneously achieve small bending radius (i.e, compact circuit) and low propagation loss have been reported [24,25]. Silicon also shows strong third-order optical nonlinearities, which can be advantageous for ultra-fast signal processing [26]. But at the same time, the material also suffers from high nonlinear loss through two-photon and free-carrier absorptions (TPA and FCA). As it is, silicon is a poor material for light sources, optical modulators, or photodetectors. But through doping, high-speed modulators and photodetectors have been demonstrated [23].

Dielectric waveguides formed by layers of silicon nitride and silicon oxide [27,28] are gaining popularity due to the potential of ultra-low loss operation. Depending on the deposition methods of these layers, post-fabrication processing, and geometry, silicon nitride waveguides can be versatilely tailored to exhibit ultra-low propagation loss (0.095-0.2 dB/cm) [29, 30], relatively compact (50-150 μm bending radius), or to have enough thickness for dispersion engineering necessary for third-order nonlinear process [30-33]. As it is, the nonlinear coefficient of silicon nitride is about 10 times lower compared to silicon. But the low loss and, crucially, the absence of TPA and FCA, make it a material of choice for microresonator based Kerr frequency combs [34,35].

**Hybrid integration and emerging materials**

It has been clear for some time for researchers in the field that none of the major material platforms discussed above can give all the required performance for integrated MWP by its own. Attempts to create an all-integrated MWP chip in a single platform (in this case InP) for functionalities like tunable filters and interference cancellation circuits have been reported with encouraging results, but issues related to waveguide loss and elevated noise due to on-chip amplifiers limit the performance of such circuits. Hence, researchers turned to a different approach, namely hybrid or heterogeneous integration, to combine different material platforms and take advantage of their individual strengths. Different approaches have been proposed (Fig 1e-g) for this, namely, chip-to-chip (hybrid) integration through vertical or edge coupling [36], or wafer-scale (heterogeneous) integration techniques such as wafer bonding or direct epitaxial growth [37, 38] which cater more mass-produced applications.

The most rapid advances occur in the area of III-V integration with silicon or silicon nitride. These circuits were aimed to provide reliable light sources and modulators (in III-V semiconductor materials) to low-loss circuits (in case of silicon nitride) or to a versatile platform with electronic integration potential (silicon). Hybrid devices have been exploited to show basic and advanced functionalities, such hybrid metal-oxide-semiconductor (MOS) Mach-Zehnder Modulators [39,40], high gain and saturation semiconductor optical amplifiers (SOAs), integrated optical and RF sources [36, 41]. Integration of multiple materials (silica, silicon nitride, III-V on silicon) was recently attempted to create a precise optical frequency synthesizer [42] (Fig 1h). Despite these advances, consensus has not been reached on which is the most suitable approach for MWP applications since, ideally, simultaneous photonic, RF and CMOS electronic compatibility [43] should be targeted.

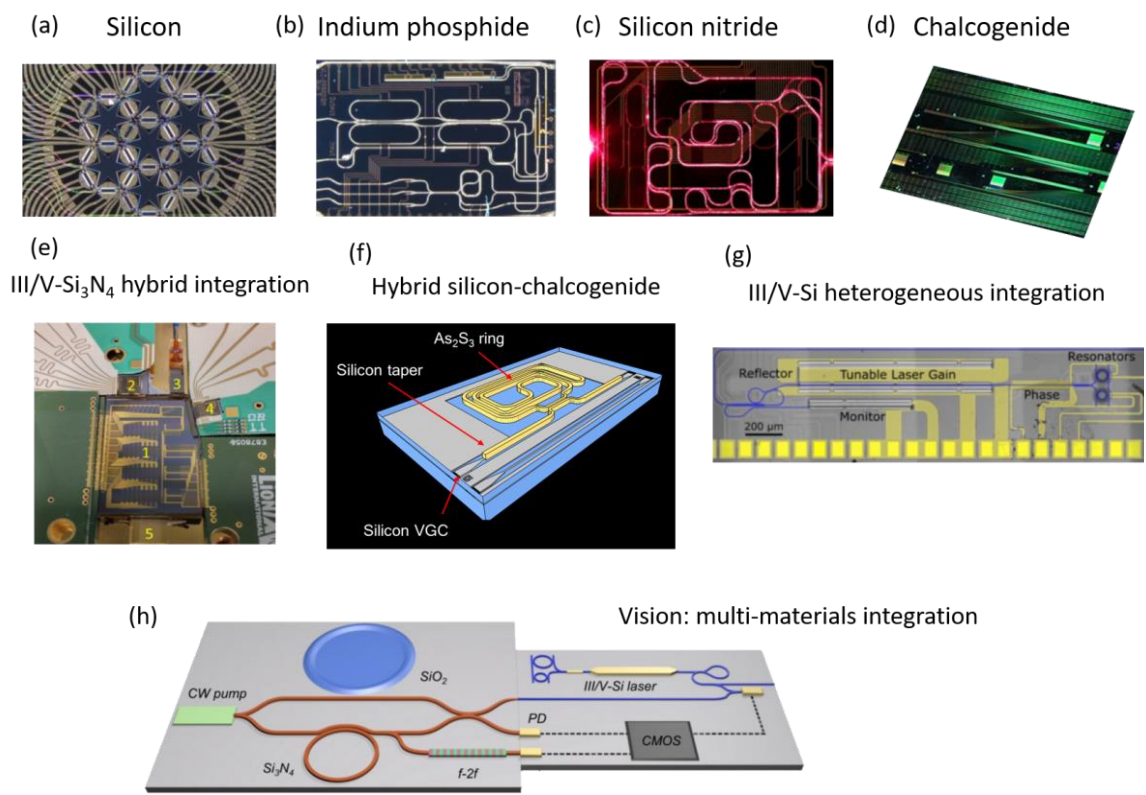

**Fig. 1. Material platforms and integration methods in modern integrated microwave photonics.** Examples of integrated microwave photonics circuits in the three major material platforms in the field for the past 10 years **(a-c)**: **(a)** programmable signal processor in silicon-on-insulator (SOI) (from [98]), **(b)** all-integrated tunable filter in indium phosphide (from [20]), and **(c)** high spectral resolution bandpass filter silicon nitride (from [36]). The emergence of nonlinear integrated MWP saw new materials explored in addition to the major platforms, including chalcogenide glasses **(d)** with enhanced opto-acoustic interactions for stimulated Brillouin scattering-based signa processing (from [81]). Recent examples of hybrid and heterogeneous integrated circuits explored for microwave photonics **(e-h)**: **(e)** hybrid integration of optical beamforming circuit in silicon nitride and modulator array and photodetector on indium phosphide (from [36]), **(f)** hybrid silicon-chalcogenide ring resonator for narrow-linewidth Brillouin laser (from [86]), **(g)** heterogeneous III/V-silicon narrow-linewidth tunable laser source with indium phosphide gain section and silicon ring resonator reflector (from [38]), (h) a vision of multi-material integration for precise optical frequency synthesis with a circuit including two optical frequency comb sources in silica and silicon nitride microresonators, and a III/V-silicon tunable laser and photodetectors (from [42])

Apart from the major platforms discussed above, other material systems have been considered as well for integrated MWP implementation, albeit in a smaller scale and volume, and were focused more on particular light-matter interactions such as nonlinear optics, opto-mechanics, and plasmonics. Chalcogenide glasses [44] (Fig 1d), a highly nonlinear material with low TPA have mainly been exploited for nonlinear opto-acoustic processing based on Brillouin scattering. Emerging materials such as Hydex [28], $Ta_2O_5$ [45], and aluminium nitride [46] can also be of interest for integrated MWP. Moreover, the availability of lithium niobate on insulator (LNOI) [47] platform has rekindled interests in lithium niobate circuits that have been shown to exhibit ultra-low loss waveguides and compact modulators, which we will be revisited when we discuss advances in optical modulators. Finally, 2D materials and very especially graphene on SOI have been proposed for the implementation of high-speed modulators [48], phase shifters, true time delay units and tunable filters [49].

**Key functionalities**

Functionalities implemented in integrated MWP span the optical modulation, generation, processing, and measurements of microwave signals. Integration not only brings advantages to these functions but also critical enhanced performance. Here we look at these categories and highlight the new technological tools available in the last few years that potentially redefine the field of integrated MWP.

**Advanced optical modulation**

Optical modulation is the first step in all microwave photonic systems, where the RF signal is encoded in the optical domain. This is pivotal step that often determines the overall system performance, including bandwidth, system loss, linearity, and dynamic range [5]. To serve the demands for higher capacity, optical modulators for modern data communications should feature high-speed and linear operation, small footprint, low loss, and high efficiency. For decades, the material of choice for optical modulation has been lithium niobate (LN) [50]. Nevertheless, the typical electro-optic LN modulators are formed either by proton exchange or in-diffusion of Titanium, leading to low index contrast waveguides with poor optical confinement, which directly leads to high drive voltages and large size.

But recent availability of LN on insulator films and improvement in LN etching techniques has changed this landscape dramatically. New high-contrast etched LN waveguides were recently used to form compact miniaturized LN modulators with a few hundred micrometers in length [51, 52] (Fig. 2a). With such a technique, impressive modulator performance was achieved including record-low drive voltage (1.4V) and a bandwidth of 40 GHz in a 20 mm-long modulator [52] (Fig. 2b). In another approach, known as rib loading, the thin film LN on insulator is interfaced with an optical waveguide from a different material, but with a similar refractive index as LN, such as $Ta_2O_5$, silicon, silicon nitride, or chalcogenides, effectively forming a rib waveguide [53-55]. The advantage of this approach is the ease of processing, as chalcogenides and silicon nitride are easier to etch compared to LN and can offer low propagation loss.

The proliferation of silicon photonics has pushed for new modulator technologies compatible with this technology. The most straightforward way to achieve optical modulation in silicon is to exploit free-carrier plasma dispersion effect, where changes in the electron and hole densities modify the refractive index and absorption of a silicon waveguide [56]. Modulators based on this effect have shown excellent performance including high-speed operation but the concept

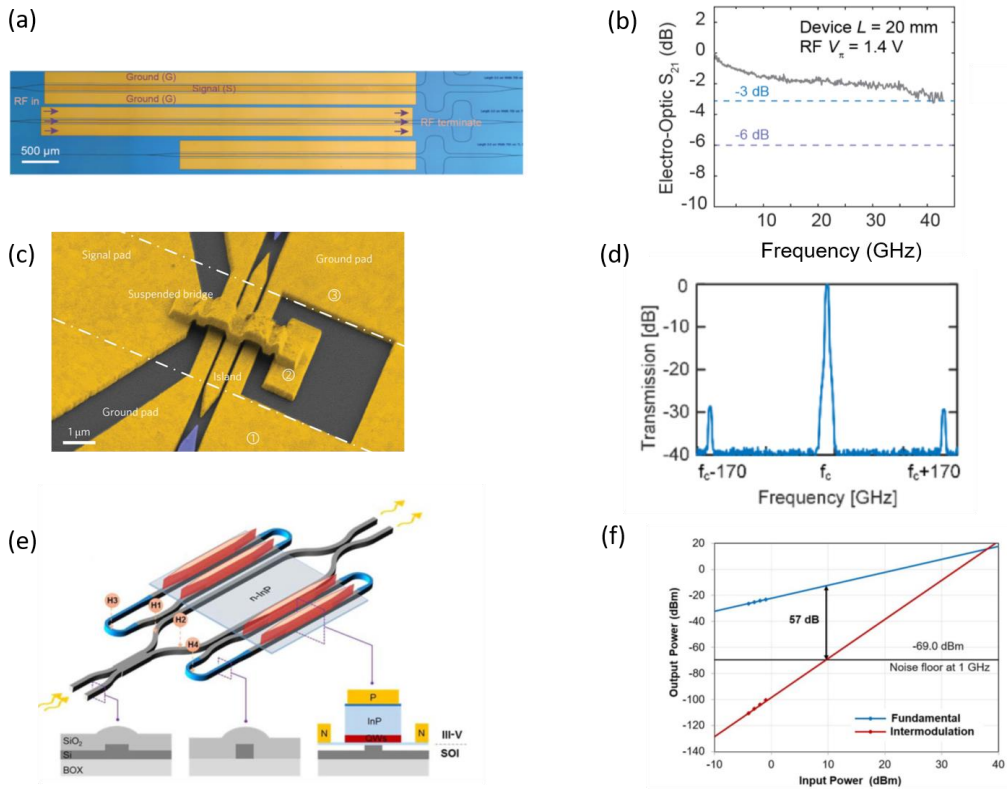

**Fig. 2. Advanced optical modulator technologies for microwave photonics. (a)** Integrated lithium niobate-on-insulator (LNOI) Mach-Zehnder optical modulator with modulator length of the order of 20 mm (from [52]), **(b)** The measured electro-optic response of the LNOI modulator. The 20mm-long device has a 3-dB bandwidth of 40 GHz and an ultra-low RF half-wave voltage ($V_{pi}$) of 1.4 V (from [52]). **(c)** Colorized SEM image of an all-plasmonic Mach-Zehnder modulator. The plasmonic interferometer is formed by the metallic island and the metallic contact pads (from [17]). **(d)** The measured optical spectrum at the output of a plasmonic modulator showing optical sidebands at 170 GHz, indicating that the modulation bandwidth up to this frequency can be achieved (from [61]). **(e)** Schematic of an ultra-linear optical modulator using a structure of heterogeneous ring-assisted Mach-Zehnder interferometer (RAMZI) modulator in III/V on silicon technology (from [65]). **(f)** The measured linearity and spurious-free dynamic range (SFDR) achieved using the ultra-linear heterogeneous RAMZI modulator. Using strong-coupled rings, SFDR of 57 dB for 1 GHz bandwidth, or 117 dB·$Hz^{2/3}$ was achieved (from [65]).

does not allow achieving the former simultaneously with low drive voltage and small footprint. For this reason, researchers proposed a hybrid integration route to combine materials having high electro-optic coefficient with strong mode confinement in silicon [57-62]. In silicon-organic hybrid (SOH) modulators [57,58], the strong confinement comes from a silicon slot waveguide, while in plasmonic-organic hybrid (POH) modulators (Fig. 2c) both optical and RF signals are guided by thin metal sheets, i.e., a metal slot waveguide where the light propagates as a surface plasmon polariton (SPP) mode [17, 59-61]. A plasmonic modulator has several advantages over an SOH modulator including ultra-short length (tens of micrometers), and higher bandwidth due to ultra-small capacitance. Bandwidth as high as 170 GHz has been achieved with plasmonic devices [61] (Fig. 2d). These modulators have also been seamlessly integrated with an antenna for direct conversion of millimetre wave to the optical domain [62]

While improvements in bandwidth, size, and energy consumption are well within reach in new class of modulator devices, an aspect that is often overlooked is linearity, which is indispensable for analog and RF photonic applications. Linearization of the silicon modulator for analog applications has been explored through heterogeneous integration with lithium niobate [63] or III-V materials [64] where RF third order nonlinearity term arising from the MZI transfer function can be cancelled by the nonlinearity of the quantum confined Stark effect in the III-V material [65] (Fig. 2e). Dynamic range as high as 117 dB.Hz$^{2/3}$ has been achieved using such modulators (Fig. 2f).

**Generation of low noise microwaves**

Low-noise and high frequency microwave sources are needed for many applications such as radar, wireless communications, software-defined radio, and modern instrumentation. Conventionally, a microwave signal is generated using an electronic oscillator with many stages of frequency doubling to generate a microwave signal with the desired frequency. The system is complicated, noisy and costly. In addition, with frequency multiplication, the phase noise performance of a microwave signal would be degraded by $10\log_{10}M^2$, where $M$ is the multiplication factor.

Among the numerous microwave generation techniques, an optoelectronic oscillator (OEO) is considered an effective solution to generate high frequency and ultra-low phase noise microwave signals. Numerous schemes to realize an OEO have been reported, but the majority are based on discrete components. To reduce the size, power consumption and cost and to increase the stability, chip-based OEOs are highly needed. Recently, Tang and co-workers have demonstrated an integrated OEO based on InP [66]. The optical components including a directly modulated laser source, a spiral shaped optical delay line, and high-speed PD were fabricated on an InP substrate. The electrical components were assembled on a PCB. Stable single-frequency oscillation at 7.3 GHz with a measured phase noise of –91 dBc/Hz at a 1-MHz offset frequency. In another approach, Zhang and co-workers have proposed an integrated frequency tunable OEO on a silicon photonic chip, in which a thermally tunable optical microdisk resonator (MDR) with a high Q factor was employed as an optical filter [67]. Thanks to the ultra-narrow notch and thermal tunability of the MDR, a microwave signal with a frequency tunable from 3 to 7 GHz and a phase noise of -80 dBc/Hz at an offset frequency of 10 kHz was demonstrated.

Another route to an integrated microwave synthesizer is through incorporation of InP-based light sources and optical amplifiers into a silicon chip. This approach combines the benefits of two different material systems to provide a unique solution to light generation and amplification. Hulme and co-workers have proposed a heterogeneously integrated silicon-III/V chip for microwave generation [41]. The chip includes a high-speed PD and two tunable laser sources the outputs of which were applied to a fast PD to generate a frequency tunable microwave signal. By tuning the wavelength of one laser source, a microwave signal with a frequency tunable from 1 to 112 GHz was generated.

On-chip nonlinear optic effect can also be employed for microwave signal generation. For example, a Kerr optical frequency comb excited in a high-Q magnesium fluoride whispering-gallery-mode (WGM) resonator was used to generate a microwave signal [68]. A stable and spectrally pure X-band microwave generation system was demonstrated, in which a microwave signal at 9.9 GHz with a phase noise as low as -60 dBc/Hz at a 10-Hz offset frequency and -120 dBc/Hz at 1-kHz offset frequency was generated. Cascaded Brillouin oscillation on an ultra-high-Q planar silica disk resonator was also used for microwave generation [69]. This microwave signal synthesizer has achieved a record low-white phase-noise floor as low as -160 dBc/Hz.

**Integrated microwave photonic filters**

Microwave filtering (MPF) is one of the most important and fundamental functions in signal processing used to separate information signals from unwanted signals such as noise and interferences. The key advantages of using photonics to implement a microwave filter include wider bandwidth, higher frequency, and larger frequency tunability. In general, there are two ways of implementing MWP filter: via a tapped-delay line architecture, where a number of RF signal copies with well-tailored amplitude and delay profile are summed to form a periodic frequency response, or via a down-converted response of an optical filter response [5, 8].

Integration of tapped-delay line filter elements has been attempted. Microresonator-based frequency combs have emerged as a powerful tool as the multi-wavelength source for tapped-delay lines signal processing [70, 71]. For example, a Kerr-based frequency comb containing 45 lines generated based on a silicon nitride chip was used as an optical source in an MPF [71]. This was the first demonstration of a single band-pass MPF employing a discrete-wavelength comb source. In [72], Metcalf and co-workers integrated a pulse shaper on an indium phosphide (InP) platform to modify the tap weight distribution using a 32-channel arrayed waveguide grating and a dedicated semiconductor optical amplifier (SOA) for line-by-line control of the channel gain. By exploiting the fast-programmable capability of the pulse shaper, a frequency tunable finite impulse response (FIR) MPF with fast reconfigurability was realized.

A fully integrated frequency-tunable MPF based on the optical filter response down-conversion concept was demonstrated on an InP platform [20]. All the components including a laser source, a modulator, an optical filter based on ring-assisted Mach-Zehnder interferometer, and a photodetector (PD), were integrated on a single chip. This is the first time that a fully integrated MPF was demonstrated. The tuning of the MPF was achieved by controlling the electrical currents into the micro-heaters to tune the optical filter. A similar filter demonstration was reported shortly after where a phase modulator, an optical filter based on micro-disk resonator, and a photodetector were all integrated in a silicon chip [73].

As an alternative to signal filtering, recently researchers from Princeton University demonstrated a monolithically integrated RF photonic self-interference cancellation system. This system, realized using InP technology, consists of a pair of lasers, a balanced photodetector, and three SOAs acting as phase and amplitude tuners. With this circuit, cancellation of unwanted RF interference up to 30 dB over the bandwidth of 200 MHz in the range of 400 MHz - 6 GHz was achieved [74].

**High resolution filtering with stimulated Brillouin scattering**

Although optical-based RF photonic filters can be tuned over a wide frequency range, the spectral resolution of most optical filters, typically in the gigahertz, are too coarse for processing RF signals where the separation between adjacent information channels can be down to only a few tens of megahertz. Breaking this barrier, a number of integrated microwave photonic filters that combine the strengths of photonic and electronic filters, namely tens of gigahertz frequency tuning with megahertz spectral resolution and an ultra-high extinction have been reported [18, 75-79]. This unique performance metrics was achieved through harnessing stimulated Brillouin scattering (SBS), a coherent interaction between optical waves and high frequency acoustic wave (hypersound), in an integrated waveguide [80-82]. Spectrally, SBS manifested in a narrowband gain resonance, shifted in frequency by about 10 GHz. In particular, SBS-based filters can exhibit linewidths of the order of 10-100 MHz. which is unmatched by most on-chip devices.

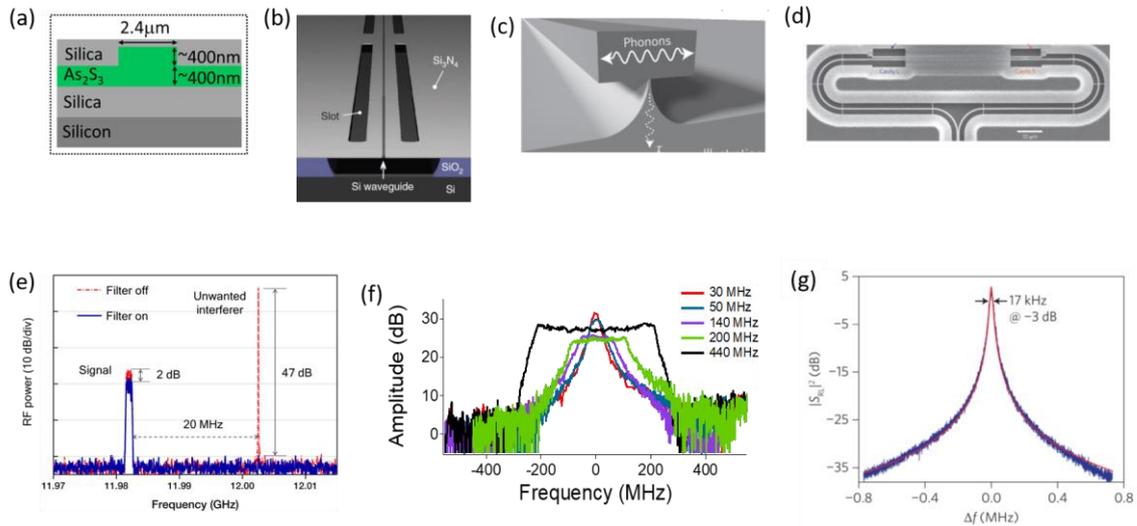

**Fig. 3. Material platforms and filtering responses of photon-phonon-based microwave photonic filters.** Structures for inducing strong stimulated-Brillouin scattering in a photonic chip **(a-c)**: **(a)** $As_2S_3$ chalcogenide waveguide. In this structure, both light and high frequency sound waves are confined on the $As_2S_3$ core. The high acoustic impedance mismatch between chalcogenides and silica allow phonon confinements (from [83]). **(b)** Hybrid silicon-silicon nitride photonic-phononic waveguide. The under-etched silicon waveguide connected to silicon nitride membranes with slits provide the necessary photon and phonon confinements. (from [82]). **(c)** Silicon pedestal waveguide partially released from the silica substrate minimizes phonon losses through the substrate (from [84]). **(d)** As alternative to Brillouin scattering, an optomechanical structure consisting of two optical microcavities connected by a phononic waveguide can act as a high-resolution microwave photonic filter (from [89]). **(e-g)** Examples of filtering experiments conducted using photonic-phononic waveguide devices. **(e)** High resolution RF photonic notch filtering based on low-power chalcogenide SBS device. The unwanted interferer 20 MHz away from the desired signal is effectively filtered without significant reduction of the desired signal power, highlighting the high resolution and rejection of the notch filter (from [18]). **(f)** Demonstration of Brilloun gain tailoring in a chalcogenide chip. By properly shaping the optical pump profile, filter with wide range tunable bandwidth and square-shaped response can be achieved (from [75]). **(g)** High resolution (17 kHz 3dB bandwidth) filtering response from the optomechanical circuit shown in (d) (from [89]).

Achieving strong SBS response in integrated waveguides requires material platforms with strong electrostriction and elasto-optic coefficients. Moreover, SBS in waveguides requires geometries with strong overlap between optical and acoustic modes. To date, the strongest on-chip SBS response was demonstrated in chalcogenide ($As_2S_3$) waveguides [83] (Fig. 3a). But achieving high gain in versatile materials like silicon is still challenging. To do so, one must combat the prohibitively high acoustic phonon leakage from silicon to the silica substrate and TPA and FCA. The groups at Ghent [84] and Yale [85] used suspended silicon structures to demonstrate 4-6 dB of SBS gain (Figs. 3b-c). An alternative approach is to explore hybrid and heterogeneous integration, where a material that supports efficient generation of SBS, for example chalcogenides, is embedded in a CMOS compatible circuit. With this approach, researchers from Australia have demonstrated 22 dB of SBS gain from a 6 cm-long $As_2S_3$ waveguide heterogeneously integrated in a silicon chip [86] (Fig. 1f).

Based on such SBS devices, high resolution and tailorable RF photonic signal processing and filtering functions have been demonstrated including high extinction, high resolution notch filters [18] (Fig. 3e), RF phase shifters and true time delay [87, 88]. Another advantage of SBS is all-optical reconfigurability. By pumping the SBS medium not only with a single laser, but with multiple lines with shaped amplitude and frequency spacing, broadened tailorable response can be achieved and one can generate unique filters with flat-top sharp-edge frequency response and tunable bandwidth [75] (Fig. 3f).

In addition to more traditional SBS devices exploiting counter-propagating pump and probe, the novel concept of phononic-photonic emitter receiver has been used to demonstrate RF photonic bandpass filter with ultra-narrow bandwidth of around 5 MHz [77]. Similarly, the device exploiting phonon routing between two opto-mechanical cavities (Fig. 3d) can be shown to generate ultra-narrow filters with impressive 17 kHz of 3-dB width [89] (Fig. 3g). These devices are promising for channel selections in spectrally crowded RF environments.

**Programmable signal processing**. Until very recently, the majority of reported integrated photonic microwave signal processors have been implemented as application specific integrated circuits (ASPICs), which are designed to optimally perform a particular MWP functionality [5,6]. This results in a lack of universality and reconfigurability for multi-functional applications. Leveraging on the strong push towards programmable photonics from related fields including quantum photonics [90-92] researchers have strived towards programmable integrated MWP devices, with two particular routes explored. First, circuits based on traditional interferometric and photonic waveguide structures are designed that incorporate the possibility of flexible programming and reconfiguration of its key parameters. Liu and co-workers demonstrated a fully reconfigurable InP photonic integrated signal processor [93], made of three active ring resonators and a bypass waveguide as a processing unit cell (Fig. 4a). These components are tuned by means of SOAs and phase modulators (PMs). With this circuit, reconfigurable signal processing functions including filtering, temporal integration, temporal differentiation and Hilbert transformation can be performed through controlling the injection currents to the SOAs and PMs. Other implementations of this approach include thermally tunable silicon nitride circuits [27] and the recently proposed programmable photonic signal processor using reconfigurable silicon Bragg gratings [94].

The second approach towards a programmable processor looks at the possibility of making a generic signal processor from a mesh of uniform tunable building blocks that can be "programmed" to support multiple functions through a software-defined operation [19, 95-98]. This concept is inspired by the concept of field-programmable gate arrays (FPGAs) in electronics and software defined networks (SDN) in telecommunications. In the two seminal demonstrations of this concept [19, 98], the tunable building block of choice is a Mach-Zehnder interferometer (MZI) composed of a tunable coupler and four input/output waveguides. The MZI can be tuned either through electro-refraction/electro-absorption effect, or by thermo-optic effect, leading to complete independent tuning of the splitting ratio and phase shift of light traversing through the building block. The MZIs can then be arranged into a two-dimensional (2D) unit cell in square, triangular, or hexagonal-type meshes (Fig. 4d). The key advantage of using MZIs as a building block is its versatile operation either as a switch or a tunable coupler providing independent tuning of the power coupling ratio and the phase shift. Interconnection of these MZIs in meshes allows feed-forward and feed-backward path selections required to synthesize many optical functionalities.

In [19], Zhuang and co-workers proposed and demonstrated a programmable optical chip connecting MZI devices in a square-shaped mesh network grid. Figure 4b shows the basic layout of this chip that was fabricated in $Si_3N_4$. It comprised two square cells and was employed to demonstrate simple FIR and infinite impulse response (IIR) impulse response filters with single and/or double input/output ports of synthetized ORRs. The processor featured a free spectral range (FSR) of 14 GHz and was programed to demonstrate a Hilbert transformer, a delay line and both notch and bandpass filters (this last functionality is shown in Figure 4c). Perez et al. then proposed triangular and hexagonal geometries [98], which bring improved performance. Impressive results have been recently reported for a waveguide mesh composed of 7 hexagonal cells fabricated in Silicon on Insulator. Figure 4e shows a photograph of the overall structure, which comprised 30 independent MZI devices and 60 thermo-optic heaters. MZI devices where independently tuned in power splitting ratio and overall phase shift by means of current injection to the heaters deposited on top of the waveguides implementing the interferometer arms. The structure is capable of implementing over 100 different circuits and functionalities. Figure 4f shows optical response of the chip when configured as an add-drop ring resonator. Several practical issues need attention to advance in this research approach. These include the tracking and stabilization of circuit parameters, the robust operation against departures of the TBUs from their designed values and their scalability and complexity. Recent publications have addressed some of these challenges [99, 100], but future research is required to address the many tradeoffs involved in the solving of these limitations.

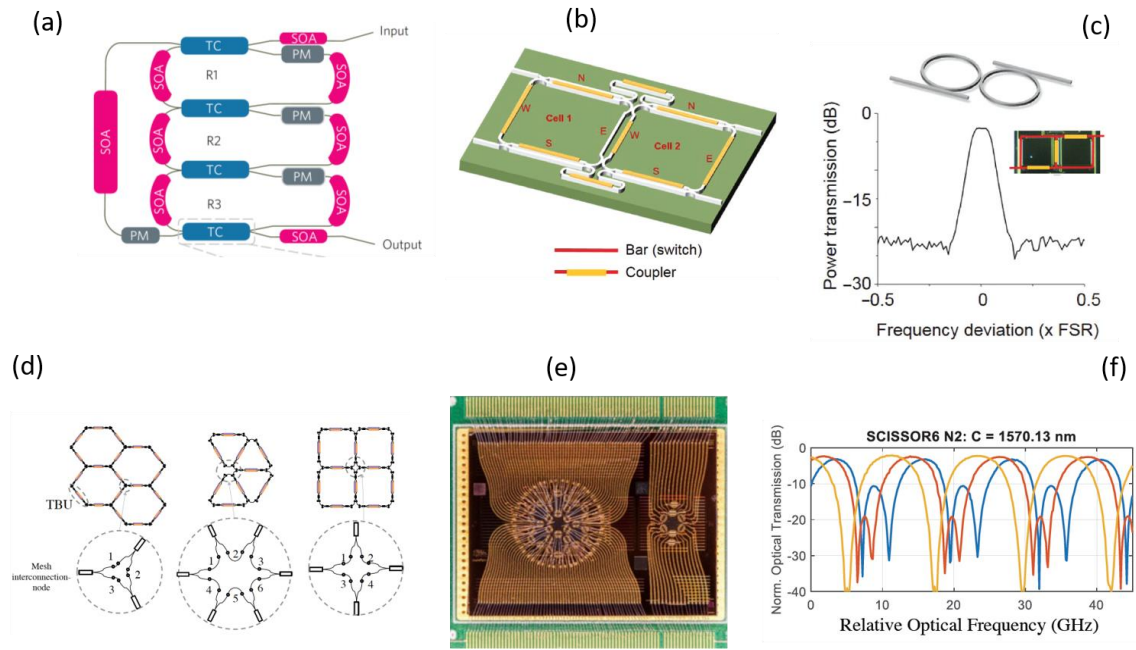

**Fig. 4. Programmable and general-purpose microwave photonic processors. (a)** The three-ring topology in indium phosphide. The device consists of three coupled-rings with phase and amplitude tunable elements and a bypass optical waveguide. The device can synthesize functions including optical differentiator and integrator as well as Rf photonic filter and Hilbert transformer (from [93]). **(b)** A programmable optical chip connecting Mach-Zehnder interferometer (MZI) devices in a square-shaped mesh network grid. The chip was fabricated using low loss silicon nitride technology (from [19]). **(c)** By thermo-optic tuning, the square-shaped mesh network can be programmed to exhibit various optical functions, including a square-shape band pass filter normally achieved using a lattice two-ring filter (from [19]). **(d)** Three general topologies of interconnected MZIs, namely triangular mesh, hexagonal mesh, and square mesh (from [97]). **(e)** The photograph of reconfigurable signal processor based on hexagonal mesh. The overall structure comprised of 30 independent MZI devices and 60 thermo-optic heaters. (from [98]). **(f)** The processor can be programmed to exhibit more than 100 distinct optical responses including an optical response from an add-drop ring resonator (from [98]).

**Current and future applications**

Thus far, the developments of integrated MWP technologies have been driven by communications-related applications such as radio-over-fiber systems. This is expected to continue with the proliferation of new concepts such as 5G systems [101,102] and the internet of things [103] (Fig. 5a). The flexible generation of pure and high frequency carriers and fast, large bandwidth beamsteering are two important concepts in the next generation radio systems with massive multiple-input multiple-output (MIMO) where integrated MWP is expected to play a key role.

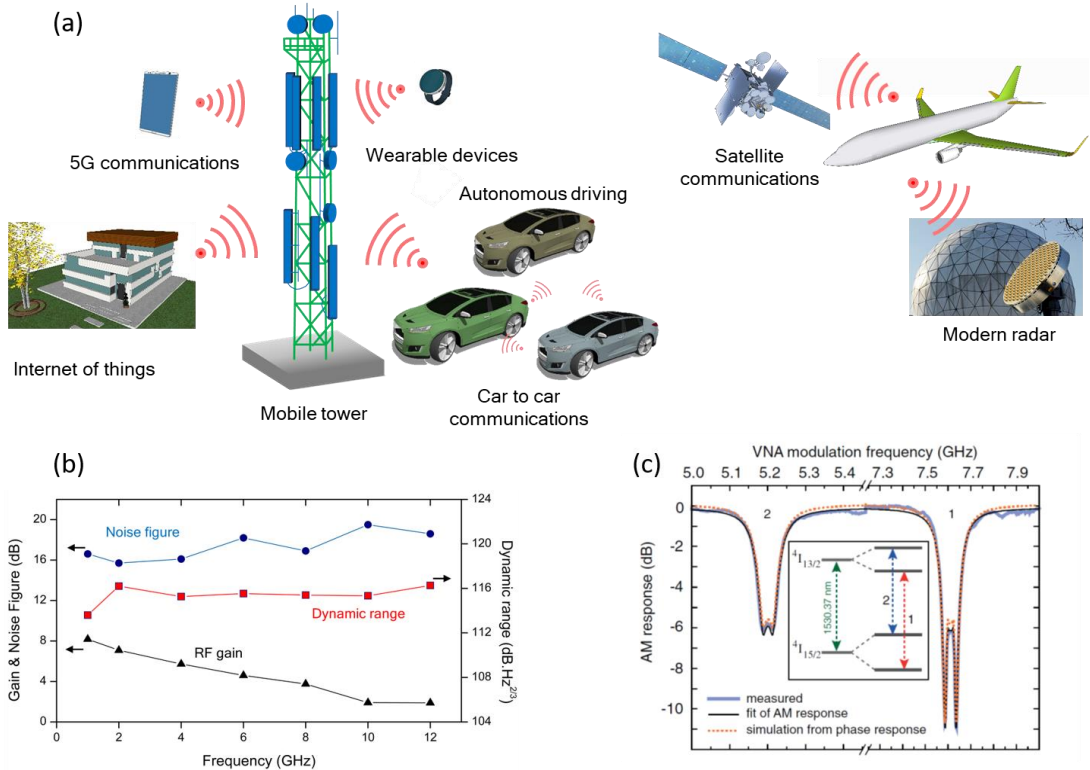

**Fig. 5. Applications and challenges of integrated microwave photonics**. **(a)** Illustration of the communication-driven applications of integrated microwave photonics, including 5 G communications, the internet-of-things (IoT), and autonomous driving applications. The advantage of size and weight of the technology open opportunities in footprint-sensitive applications such as satellite communications. The broadband advantages of the microwave photonics also enable modern wideband radar. **(b)** To serve many applications outlined in (a), integrated microwave photonic devices and subsytems should demonstrate high performance, including high gain, low noise figure, and high dynamic range. Such performance has been recently demonstrated in partially-integrated filter in silicon nitride platform (from [109]). **(c)** In addition to communication-driven applications, new implementations of microwave photonics have emerged. As an example, RF photonic analysis has been used to measure microwave transmission through Er:LYF crystal. The inset shows addressed optical levels of Er:YLF crystal and transitions corresponding to the absorption components 1 and 2 (from [106]).

Photonic-based radar systems, on the other hand, are another emerging application of this technology [9, 104]. For example, Ghelfi and co-workers have proposed a photonics-based fully digital coherent radar [9]. With the use of microwave photonic techniques, low-noise and high frequency microwave signals were generated at the transmitter, and at the receiver, photonic-assisted microwave signal processor was employed to handle wideband received signals.

The technology also finds applications in size and weight-sensitive applications such as space and satellite communications [105] (Fig. 5a). Here flexible optical filtering and beamforming circuits are developed to replace heavy waveguide-based components.

A host of new applications beyond communications have been explored too. For example, the concept of optical vector analysis, which is microwave sideband modulation and generation, was used for the spectroscopy of ultra-narrow optical transitions of isotopically purified crystals [106] (Fig. 5c). Here the RF modulation sidebands are employed to excite microwave photon (i.e very fine energy resolution) transitions on top of a two energy level transition enabled by an optical carrier frequency. The on-chip integration of this simple testing system, where a central part is reserved to host/deposit the material and/or substance to be tested opens unprecedented opportunities for high-resolution testing of biomedical and organic substances.

**Challenges to address**

To be deployed in actual RF systems, integrated MWP devices have to achieve comparable performance to RF devices. These translate into stringent requirements in terms of RF characteristics including no loss of the signal of interest (often reflected as zero or positive RF link gain in the decibel scale), a single-digit noise figure (NF), and a high spurious-free dynamic range (SFDR) of more than 120 dB.Hz$^{2/3}$ [107]. Such performance has been achieved in analog fiber-optic links for RF signal transport and distribution but in these demonstrations no RF photonic processing functionality was implemented. In contrast, the link performance of reported IMWP systems demonstrations is either way below the required target performance, or more often, not even considered or reported.

Only recently, researchers have made progress in achieving partially-integrated RF photonic functional systems simultaneously with high performance, including positive link gain, a low noise figure of 15.6 dB and a high SFDR of 116 dB.Hz$^{2/3}$ [108, 109] (Fig. 5b). Such a breakthrough in performance was achieved by putting together a number of known approaches for link optimization including low biasing of a MZ modulator while increasing the input optical power, minimizing the loss in the optical waveguide platform, and using a high power handling photodetector composed of multiple photodiodes to ensure linearity. It is thus intriguing to see how these levels of performance can be extrapolated to fully integrated systems.

**Outlook and perspectives**

Adoption of the new technological tools described earlier not only equipped the microwave photonics with advanced functionalities but also expands the field considerably to allow many intersections with other growing fields in photonics, potentially creating new concepts and paradigms. Here we give our perspectives on these new concepts.

The manipulation of phonons and high frequency sound waves in integrated devices can effectively bridge classical RF photonics processing with cavity opto-mechanics [110] with potential broadened applications in sensing to quantum information science. New architectures that manipulate phonons generated optically and transduced through radio-frequency fields hold the promise of enhanced signal processing [111] (Fig. 6a).

Both ASPICs as well as versatile integrated reconfigurable processors provide the low footprint, controllable, compact and stable environments that can enable the demonstration of new MWP concepts. For instance, multiple port chips can enable the possibility of Multiple Input/Multiple Output (MIMO) MWP opening the door to parallel linear processing [112] and space division multiplexing [113, 114]. Furthermore, analog photonic principles combined with unitary $N$x$N$ transformations have been proposed as a key technology for the implementation of spiking and reservoir neuromorphic photonic systems, deep learning (Fig. 6b) and brain inspired processing [115-117]. For example, these new concepts can be implemented in the wideband fingerprinting of crowded wireless environments for cognitive radio applications [115] (Fig. 6c). Interestingly, integrated MWP systems based on photonic ring cavity cascades modulated by RF signals have also been proposed for the implementation of a new generation of processing systems based on the physics of synthetic dimensions [118].

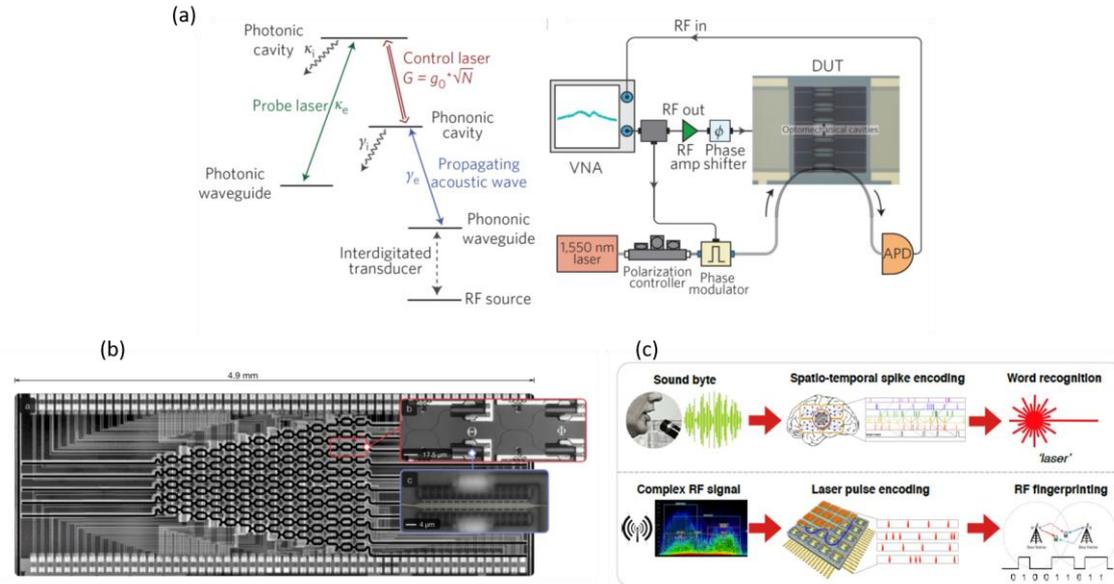

**Fig. 6. New opportunities for integrated microwave photonics**. **(a)** High frequency phonons have the ability to directly bridge radio frequency domain and the photonic domain. In recent demonstration, an optomechanical device in piezoelectric materials can couple phonons generated from RF sources and phonons generated optically. The figure on the left depicts the level diagram of such interactions. The figure on the left depicts the device and measurement setup to tailor the photon-phonon-RF interactions. This new concept opens new opportunities in advanced signal processing and tailored responses beyond only RF photonic concepts (from [111]). **(b)** Programmable optical processors intersect integrated microwave photonic fields with new emerging fields including integrated quantum photonic and brain-inspired (neuromorphic) photonics. The figure depicts the optical micrographs of programmable nanophotonic processor in silicon photonics used for deep learning applications. The chip size is 4.9 mm by 1.7 mm and is composed of 56 Mach-Zehnder interferometers, 213 phase shifters, and 112 directional couplers (from [116]). **(c)** Illustration of neuromorphic photonic concepts implemented for RF fingerprinting of complex and crowded RF environments for cognitive radio applications. Applying operations at the front-end of RF transceivers could offload complex signal processing operations to a photonic chip, and address bandwidth and latency limitations of current digital signal processing (DSP) solutions (from [115]).

The concept of parity-time symmetry employed in laser optics to achieve single-mode lasing [119-121] has recently been employed in microwave photonics systems for filterless single-frequency microwave generation. For example, an opto-electronic oscillator based on parity-time symmetry was demonstrated to generate a single-frequency and ultra-low phase noise microwave signal without using a microwave or optical filter [122], which overcomes the long existing mode-competition and mode-selection challenge which has severely limited the development and wide applications of opto-electronic oscillators. The employment of parity-time symmetry will also ease photonic integration of OEOs since an ultra-narrow band microwave or optical filter is hard to implement in an integrated chip.

These new concepts in integrated MWP driven by the new understanding in device physics and the tremendous growth of integrated photonics represent a new phase in the field beyond the pivotal first phase driven by telecom-related applications. Such new explorations will sustain the growth of these fields and bring new exciting opportunities.